\documentclass[letterpaper]{article} 
\usepackage[]{aaai25}  
\usepackage{times}  
\usepackage{helvet}  
\usepackage{courier}  
\usepackage[hyphens]{url}  
\usepackage{graphicx} 
\usepackage{natbib}  
\usepackage{caption} 
\usepackage{algorithm}
\usepackage{algorithmic}

\usepackage{newfloat}
\usepackage{listings}
\DeclareCaptionStyle{ruled}{labelfont=normalfont,labelsep=colon,strut=off} 
\floatstyle{ruled}
\newfloat{listing}{tb}{lst}{}
\floatname{listing}{Listing}
\title{Evaluating Intra-firm LLM Alignment Strategies in Business Contexts}
\author{
    Noah Broestl\textsuperscript{\rm 1}\equalcontrib,
    Benjamin Lange\textsuperscript{\rm 2\rm 3}\equalcontrib,
    Cristina Voinea\textsuperscript{\rm 4}\equalcontrib,
    Geoff Keeling\textsuperscript{\rm 5}\equalcontrib,
    Rachael Lam\textsuperscript{\rm 1}
}
\affiliations{
    \textsuperscript{\rm 1}Boston Consulting Group\\
    \textsuperscript{\rm 2}Ludwig-Maximilians-Universität München\\
    \textsuperscript{\rm 3}Munich Center for Machine Learning\\
    \textsuperscript{\rm 4}Uehiro Oxford Institute, University of Oxford\\
    \textsuperscript{\rm 5}Google, Paradigms of Intelligence Team\\
    broestl.noah@bcg.com, benjamin.lange@lmu.de, cristina.voinea@uehiro.ox.ac.uk,
    gkeeling@google.com,
    lam.rachael@bcg.com
}

\begin{document}

\maketitle

\begin{abstract}
Instruction-tuned Large Language Models (LLMs) are increasingly deployed as AI Assistants in firms for support in cognitive tasks. These AI assistants carry embedded perspectives which influence factors across the firm including decision-making, collaboration, and organizational culture. This paper argues that firms must align the perspectives of these AI Assistants intentionally with their objectives and values, framing alignment as a strategic and ethical imperative crucial for maintaining control over firm culture and intra-firm moral norms. The paper highlights how AI perspectives arise from biases in training data and the fine-tuning objectives of developers, and discusses their impact and ethical significance, foregrounding ethical concerns like automation bias and reduced critical thinking.  Drawing on normative business ethics, particularly non-reductionist views of professional relationships, three distinct alignment strategies are proposed: supportive (reinforcing the firm's mission), adversarial (stress-testing ideas), and diverse (broadening moral horizons by incorporating multiple stakeholder views). The ethical trade-offs of each strategy and their implications for manager-employee and employee-employee relationships are analyzed, alongside the potential to shape the culture and moral fabric of the firm.
\end{abstract}

\section{Introduction}

Instruction-tuned Large Language Models (LLMs) are increasingly deployed as general-purpose AI Assistants in firms.\footnote{The deployment of LLMs as general-purpose AI assistants is likely to apply more broadly to non-commercial organizations, but for the purposes of this paper we focus on AI Assistants deployed in firms.} Generally, these systems are enterprise deployments of products like ChatGPT. These systems are used in tasks that could be broadly considered ``thought partnership,'' completing cognitive tasks that were previously completed by individuals or groups of employees acting in collaboration. AI Assistants are distinct from the use of Generative AI as a component of applications produced by the firms; for example, a non-AI Assistant application could be the integration of Generative AI into a tool that summaries unstructured data for compliance purposes. Enterprise deployments of ChatGPT-like products are generally provided out of the box with little to no modification. They are typically available to every member of the firm, and they are intended to increase productivity by supporting common use cases such as conducting research, editing documents, and brainstorming. These uses of LLMs allow employees to offload some of their cognitive tasks from drafting emails to strategic thinking.

However, the impacts of AI assistants in firms are not limited to the output of individual content. Rather, AI assistants will shape the way employees interact and the culture of the entire firm as these impacts happen at scale. AI Assistants are now deeply embedded into workflows, so they have a strong influence on how decisions are made, and how colleagues relate to one another within a firm. But, as we show in this paper, AI Assistants embed distinct ``perspectives''---that is, they exhibit behavioral dispositions which seep into decision-making processes, ultimately influencing how authority is exercised, how collaboration happens, and how employees and managers understand their roles, obligations, and relate to one another. 

This paper argues that firms are morally required to be intentional about how AI Assistants are developed and deployed. Alignment is a strategic and ethical imperative if and to the extent that firms want to retain control over their cultures\cite{32}. Alignment, as we understand it, is not only a technical matter, but a fundamentally ethical and relational one. To that end, we argue for the need to understand the alignment of the perspectives of AI Assistants in relation to the objectives and values of firms. We propose three distinct strategies (supportive, adversarial, and diverse) for the deployment of AI Assistants and analyze the ethical trade-offs that stem from each strategy. We then show that the choice of strategy impacts intra-firm relational dynamics and thus requires an evaluation of how the specific norms governing manager-employee and employee-employee relationships will be shaped by the introduction of AI Assistants.

The paper proceeds as follows. Section \ref{sec:perspective} develops a conception of what it means for an AI Assistant to have a perspective. Section \ref{sec:impact} discusses where these perspectives may have an impact on the firms. Section \ref{sec:norms} draws on scholarship in normative business ethics, to better characterize  the intra-firm moral norms and significant intra-firm human relationships that AI Assistants are likely to affect. Against this backdrop, Section \ref{sec:alignment} then develops three AI Assistant alignment strategies and analyzes their ethical trade-offs. Section \ref{sec:conclusion} concludes the discussion. 

\section{Perspectives of AI Assistants}
\label{sec:perspective}

We can think of AI Assistants as having a perspective on the world. This perspective can be understood as a set of behavioral dispositions to address user queries in line with particular social, political, ethical or cultural biases. For example, an AI Assistant used in a political consultancy firm could be disposed to answer questions in line with neoconservative commitments. Furthermore, firms may find certain model perspectives more desirable than others—in this case, a political consultancy firm may want an AI Assistant that is disposed to answer questions in line with neoconservative commitments.

Metaphysically, perspectives can be interpreted in two ways \cite{1}. On a causal reading, an AI Assistant has some unique perspective that features in causal explanations of its behavior in roughly the same way that a human’s perspective might causally explain their behavior (e.g. Anne advised Beth to $\varphi$ because Anne is a neocon, and the AI Assistant advised Beth to $\varphi$ because the AI Assistant is a neocon). On an epiphenomenal reading, the perspective of an AI Assistant is not a cause of the AI Assistant’s behavior, but rather something that is read onto the AI Assistant’s behavior to interpret that behavior in a human-like way. On this latter picture, the AI Assistant begins a dialogue in a superposition over the set of possible perspectives, and this superposition narrows as the dialogue advances and additional consistency requirements are imposed on what perspectives the AI assistant can reasonably be interpreted as occupying. For example, if the AI Assistant outputs text that is out-of-line with neocon commitments, then it can no longer be interpreted as having a neocon perspective.

Here we interpret perspectives in line with the causal reading. Perspectives, so understood, result primarily from the interplay of two classes of factors.\footnote{It may be the case that additional factors causally impact the perspective realized in LLMs; for example, model architecture decisions, hyperparameters used in training, and data pre-processing techniques \cite{2} Even so, the training distribution and Reinforcement Learning with Human Feedback (RLHF) objectives are plausibly the principal determinants of a model’s perspective in the minimal sense that these factors offer two of the most straightforward intervention points for explicitly engineering a perspective into the model.} The first class of factors is biases encoded in the LLM that undergirds the AI Assistant. Biases in the distribution of pre-training data or in the pre-processing of pre-training data can result in LLMs being disposed to answer user queries in line with particular sociopolitical points of view \cite{2,3}. For example, differential representation of Western versus non-Western text data in the pre-training corpus can result in LLMs having more granular representations of Western concepts which can dispose LLMs to answer queries in line with Western values \cite{4,5,6}. The second class of factors concern developer objectives when fine-tuning LLMs via supervised fine-tuning, reinforcement learning from human feedback, and other related methods \cite{7,8,9}. Here developers can explicitly fine-tune LLMs to reflect particular value commitments such as helpfulness, honesty and harmlessness \cite{10,11,12}. Insofar as LLMs possess explicit representations of sociopolitical perspectives \cite{13}, it is possible to fine-tune LLMs to answer user queries from the relevant sociopolitical perspectives. 

AI Assistant perspectives, so understood, can very along at least two dimensions. First, perspectives can be more or less \textit{specific}. In particular, at one extreme are schematic perspectives such as the AI Assistant having a broadly right-or-left-leaning bias; and at the other extreme are highly specific perspectives such as the AI Assistant reflecting the ideological outlook of Noam Chomsky or Henry Kissinger. Second, perspectives can be more or less \textit{consistent}. In particular, certain perspectives---such as those deliberately engineered into the underlying LLM using supervised fine-tuning or Reinforcement Learning with Human Feedback (RLHF)---are likely to resonate persistently through the AI Assistant's answers in response to a broad class of queries. These perspectives are consistent in that they consistently explain relevant features of the content of the AI Assistant's answers. Conversely, other perspectives---such as those resulting from biases in the training distribution that pertain to particular topics---may be less consistent insofar as these perspectives explain the content of the AI Assistant's answers in a minority of cases that relate to the relevant topics. 

In many cases, firms deploying AI Assistants and the employees at those firms will have little if any agency over what perspectives AI assistants used by their employees exhibit. On the one hand, as most employees have low to no technical expertise, their primary interaction with AI Assistants will occur through a chat-based interface, which does not expose model configuration or tuning options. Fine-tuning capabilities are only available through API integrations, inaccessible to and not readily usable by non-technical users. On the other hand, AI Assistants may be pre-trained and fine-tuned by independent developers which provide minimal scope for firms to shape the perspective of the assistant. Here the perspective of the AI Assistant may be unknown to the firm, and potentially unknown to the developer also—if, for example, the perspective of the AI Assistants has not been defined explicitly in the model specification or the developer has not attempted to evaluate what perspective is implicit in the model. In other cases, firms will have greater agency over the perspective exhibited by AI Assistants. While most firms will lack the ability to pre-train the underlying LLM, developers may afford them the option to fine-tune LLMs in line with a desired set of values or commitments. Overall, it matters that firms are intentional about the perspective of the AI Assistant used, either by making consumer decisions in line with relevant values or by fine-tuning the underlying LLMs to ensure alignment with those values. Else, firms risk having perspectives imposed upon them with potentially broad impacts.

\section{The Impact of AI Assistant Perspectives}
\label{sec:impact}

The perspectives that AI assistants bring into firms are likely to have wide-ranging impacts. AI Assistants are not merely tools. Rather, they are active participants in the creation of content ranging from mundane emails through to strategic vision documents. The fact that these systems imbue a perspective and participate in the broad generation of content renders plausible the idea that that the systems' perspectives will shape the content that they contribute to in much the same way that the perspectives of individual employees or team perspectives influence the output of traditional collaboration without AI Assistants.

Indeed, the impact of AI Assistant perspectives might be more significant than the impact the perspective of an individual employee. This is indicated by research on the topics of automation bias \cite{25} and reduced critical thinking in interactions with AI Assistants \cite{18, 26}. The evidence suggests that users often place undue trust in information generated by automated systems, sometimes at the expense of their own judgment. The impact of this misplaced trust is that the perspectives presented by the AI Assistant may not be sufficiently challenged by the humans who are reviewing the output. Additionally, a specific risk of these particular technologies is how they may directly impact critical thinking in humans who are using the system \cite{18, 26}. This combination of over-reliance and diminished critical engagement makes it plausible that the perspective that is injected into the content which AI Assistant contributes to is less likely to be scrutinized, amplifying their overall impact on decision-making and understanding.

Reduced critical assessment of AI Assistants' outputs, alongside increased trust on them, means that people will challenge them less frequently. An implication of this phenomenon in business settings is that employees and teams will defer to AI Assistants more and more, even within collaborative interactions. Consequently, AI Assistants will influence not only specific decisions and outcomes, but also the interpersonal relationships and communication patterns that lead to those outcomes, reshaping the culture of the firm. In what follows, we show why the focus should be not only on how AI Assistants shape specific outcomes, but also on how they shape moral norms that govern professional conduct \textit{within} firms.

\section{AI Assistants and Intra-firm Norms}
\label{sec:norms}

The previous section argued that AI Assistants, when deployed as general-purpose tools, express perspectives that shape how users think, deliberate, and act. These perspectives can influence not only individual decisions, but also the relationships between employees, or more broadly, a firm’s culture. The extent of the influence that AI assistants could have over a firm's culture makes it plausible that intra-firm alignment is not only a technical issue, but also a strategic one for firms. 

To better understand the ethical implications of AI Assistant perspectives and their influence over firms, we suggest orienting the ethical analysis around the moral norms that govern professional conduct \textit{within} firms.\footnote{Note that we are here primarily concerned with intra-firm normativity as opposed to other social or psychological considerations.} This orientation entails that rather than treating AI Assistants as neutral interfaces, it is necessary to consider how they interact with, reinforce, or disrupt moral relationships that govern intra-firm moral conduct between managers and employees.

Normative business ethics provides a foundation for analyzing these dynamics. Normative business ethics concerns the moral responsibilities that exist in the context of a firm \cite{30}. While business ethics is often concerned with the \textit{external} responsibilities of firms such as duties to consumers, communities, or the environment \cite{35,27,28}, there is a parallel domain concerned with \textit{internal} moral obligations \cite{31, 29}. Intra-firm business ethics thus concerns the moral principles and standards that guide behaviour and decision-making within a company; specifically the relationships between its employees, managers, departments, and leadership. Accordingly, intra-firm business ethics is concerned with questions such as fair treatment, solidarity, and respect of employees and among colleagues, fulfilling one's duties of professionalism and loyalty, abiding by the code of conduct, values, and principles, and purpose. To illustrate: A senior engineer has different ethical responsibilities than a junior analyst, not only by virtue of their role in the firm, but because of their decision-making authority and the implications of their actions. The same point applies to managers, who must balance the strategic priorities of the firm with attentiveness to the needs and well-being of employees. 

If AI Assistants are embedded in internal decision-making processes and workflows across different hierarchies, they will forseeably affect the moral dynamics of a firm or a firm's culture---the shared values, beliefs, assumptions, and behaviors that characterize how employees interact, make decisions, and conduct themselves \cite{32}. AI assistants may assist a manager in structuring deadlines, supporting an employee in drafting a report, or mediating between departments with different priorities. In doing so, they affect how intra-firm obligations are perceived and fulfilled. For this reason, AI Assistants should not be understood merely as tools that support productivity. Their presence changes how employees make decisions, how authority is exercised, and how roles are interpreted. The moral significance of AI Assistants thus arises precisely because they influence the norms that define how colleagues, subordinates, and superiors relate to one another.

To evaluate what kinds of AI assistant alignment are appropriate, we can consider how different accounts  of intra-firm business ethics characterize the moral duties within a firm. These theories offer complementary and, in part, competing views on the intra-firm moral responsibilities that exist within hierarchically structured firms. These considerations will in turn position us to derive strategies of intra-firm alignment of AI Assistants in the subsequent section. The strategy is therefore to examine the degree to which AI Assistant alignment can be construed in line with these different conceptions and to determine whether concrete forms of operationalization can be derived. 

There are two main ways to conceptualize intra-firm duties. One  view is \textit{reductionist}. On this account, moral duties within firms are best understood as constraints designed to prevent moral hazard in principal-agent relationships \cite{29,31}. These constraints seek to limit the risk that employees  act against the interests of managers, whether through negligence, conflict of interest, or opportunism. Under a reductionist account, the goal of AI Assistant alignment is primarily to support the prevention of employee misconduct or misjudgment. AI Assistants would hence act as guardrails that enforce compliance with existing normative expectations. This view supports a narrow understanding of intra-firm alignment, focused on avoiding harm rather than promoting positive values such as collegiality, moral courage, or critical reflection.

By contrast, \textit{non-reductionist} approaches maintain that intra-firm moral norms are \textit{sui generis} \cite{33,34}. Rather than treating relationships as contractual mechanisms for minimizing risk, they emphasizes the role-specific responsibilities that arise between individuals within a firm depending on their relationships \textit{qua} their roles within the firm. A manager may owe duties of fairness, support, and guidance to subordinates, for example; and colleagues may owe one another solidarity, constructive feedback, and assistance \cite{36}. From this point of view, AI Assistants should not simply be aligned to reduce risks, but to support the moral quality of professional relationships. AI Assistants may accordingly play a role in helping employees uphold professional standards, resolve tensions between obligations, or engage more thoughtfully in difficult decisions.

\section{Alignment Strategies}
\label{sec:alignment}

Non/mis-aligned AI Assistants could erode a firm’s control over decision-making processes, while also shaping a firm's overall culture in unintended respects. Accordingly, as we previously argued, firms must be intentional about aligning AI Assistants. Non-reductionist approaches to intra-firm business ethics offer an appropriate way for thinking of how alignment should be done, as it emphasizes how AI Assistants influence the moral fabric of workplace relationships. Plausibly, firms should evaluate and shape their AI Assistant’s perspective so that its presence consistently reflects and reinforces the relational norms, role expectations, and power dynamics appropriate to the firm’s workflow.  This is consistent with recent research calling for AI chatbots designed to reflect specific socio-functional roles and human-AI relationships \cite{14}. 

In what follows, we advance three forms of intra-firm alignment: 

\begin{enumerate}
    \item \textbf{Supportive}, where the AI Assistant reinforces the firm's declared mission;
    \item \textbf{Adversarial}, where the AI Assistant helps stress-test a firm's mission;
    \item \textbf{Diverse}, where the AI Assistant helps broaden a firm's moral horizon by pointing out additional stakeholder perspectives.\footnote{For our purposes, we are setting aside the possibility of what one might call ``neutral'' forms of alignment, whereby a hypothetical AI Assistant's perspective is neither supportive or adversarial across all potential issues out-of-the-box and remains in that state.}
\end{enumerate}

Each alignment strategy shapes everyday interactions within a company, which means that it has an impact on the firm's moral trajectory from the inside out.

\subsection{Supportive}

Supportive AI Assistant reflect a business’s values, goals, and strategies, as suggested by \cite{15}. This alignment strategy involves fine-tuning the underlying LLM on company-specific data or guidelines, ensuring that its outputs are consistent with the firm’s mission, decision-making norms, and ethical code.

\subsubsection{Manager-Employee Relationship:}

The overall goal of the supportive alignment strategy is to promote the company’s strategic direction while supporting ethical leadership practices. A supportive AI Assistant should assist managers in providing justifications and decisions that uphold corporate values, while also encouraging consideration of employee feedback and well-being. For employees, the aligned AI Assistant could serve as a cognitive aid, helping them fulfill tasks in ways that align with managerial expectations and the broader culture of the firm.

\begin{itemize}
    \item\textbf{Example:} In a company prioritizing innovation and speed, the AI Assistant might help managers structure deadlines and incentives to maximize fast delivery, but should also suggest balancing this with employee burnout risks and overall professional well-being. Also, employees might use an AI Assistants to offload parts of their work, such as brainstorming ideas or finishing up a task, in ways that align with managerial expectations and the desired culture of the firm.
    \item\textbf{Ethical Issues:} Already, it is documented that LLM-based agents exhibit sycophantic behaviors, flattering users and telling them what they want to hear \cite{16, 17}. If the leadership’s goals are ethically flawed (e.g., prioritizing speed at the cost of user safety), the AI Assistant may simply reinforce those flaws instead of questioning them. The problem then is that unlike managers, who can be questioned or criticized by peers and subordinates, AI Assistants can subtly reinforce power asymmetries by making the dominant perspective more salient, appear more justified, or reasonable. 
    
    In addition, AI Assistants may prioritize managerial views and the company’s overall goals, so that employees feel less empowered to raise ethical concerns if AI Assistants consistently validate leadership decisions. If an AI Assistant is aligned with a business strategy and values, and if the employees know this, AI assistants might end up encouraging over-reliance on them without sufficient scrutiny. Already, it has been shown that workers relying on LLMs tend to lose their critical thinking skills and confidence, which can lead to deskilling, be it professional or moral \cite{18}. Ethically, this means important decisions might be made on AI advice that no one double-checks, especially if it aligns with what shareholders want to hear.
\end{itemize}

\subsubsection{Employee-Employee Relationship:}

A supportive AI Assistant would promote or help foster the declared shared values of a firm, as captured by a code of conduct, mission statement, etc. In addition, supportive AI Assistants could help strengthen solidarity between employees, which might include encouraging constructive feedback, supporting a colleague through an unforeseen workload, or facilitating mutual problem-solving. 

\begin{itemize}
    \item\textbf{Example:} A supportive AI Assistant could help employees develop strategies that advance the team’s overall performance rather than merely advancing individual performance. Or a supportive AI Assistant could help fulfill collegial duties more generally; for example, by helping employees provide feedback constructively or nudging them to help each other out. Equally, a supportive AI Assistant could ease inter-departmental collaboration. Take the case of a computer science research company that values ethical research. Here, a supportive AI Assistants could nudge engineers to respect internal publication/ethical norms that promote rigor over quick results.
    \item\textbf{Ethical Issues:} Reinforcing or promoting a stated ethos can suppress valuable dissent and alternative problem-solving approaches. Moreover, employees with minority perspectives may feel marginalized if the AI Assistant only supports dominant team values. Existing departmental or professional biases (such as engineering over ethics) may be amplified without space for contestation. Another issue is the potential reinforcement of an undesirable static culture in a firm that makes more fluid dynamics such as growth or the pursuit of creativity less likely. By constantly echoing back to employees' current values and practices, the AI Assistant could make it harder for the firm to evolve, experiment with new ideas, or respond to emerging challenges \cite{18}. So what would be aimed at supporting cohesion and collegiality may come at the cost of flexibility and innovation.
\end{itemize}

\subsection{Adversarial}

Adversarial models are supposed to help managers and employees stress-test and uncover weaknesses in their ideas, while also forcing justification of decisions and pointing out ethical issues or biases that the users/teams may have overlooked (as suggested here \cite{19}). Unlike the aligned model, an adversarial AI agent can critically evaluate the user’s viewpoint, potentially leading to better-informed decisions and enhancing critical thinking. This strategy involves fine-tuning the model on examples where a firm's values/norms are in tension (for example, business ethics case studies), as well as on a whole host of examples of how to critically assess constructively arguments. For instance, a dataset might include dialogues where a proposed product launch is questioned on privacy grounds, despite alignment with company speed goals. Also, the model could be conditioned to take the role of a ‘socratic’ colleague, who questions assumptions and ideas \cite{20}. 

To clarify, by ‘adversarial’ we do not mean that the AI Assistant should by default oppose the firm’s values. Rather, the adversarial AI Assistant would operate within the space of those values to challenge how they are interpreted or applied. Take the case of a company that values "brutal honesty"; an adversarial AI Assistant wouldn’t encourage the opposite of honesty, that is, deceptiveness, but could challenge whether a decision really reflects this value or whether self-serving motives are disguised as honesty. In this sense, adversarial alignment is not the negation of values but a method of critical engagement with them.

\subsubsection{Manager-Employee Relationship:}

An adversarial AI Assistant can support managers by drawing attention to risks, stakeholder concerns, or broader obligations, like legal compliance, user safety, and employee well-being that are not explicitly stated in an ethical code or do not transpire in the culture of a firm, rather than simply reinforcing immediate business objectives. For employees, adversarial AI Assistants could help by surfacing concerns that might otherwise be difficult to raise in hierarchical settings, thereby enabling more inclusive decision-making.

\begin{itemize}
    \item\textbf{Example:} An adversarial AI Assistant could challenge a manager’s push for a rushed product launch by pointing out risks to user privacy, regulatory non-compliance or even impact on employee’s well-being (something along the lines of ``subordinates already have other deadlines for projects X and Y, consider how adding another one will affect them''). In a company where loyalty is prized, adversarial AI Assistants could nudge managers to consider whether blind loyalty might endanger long-term reputation. It could also help employees challenge managers’ decisions by voicing concerns---as if from a third-party perspective---in the employee’s stead.
    \item\textbf{Ethical Issues:} Adversarial AI Assistants might contribute to the erosion of managerial skills. If managers rely too heavily on the AI’s adversarial feedback, they risk becoming overly deferential to the AI Assistant’s objections, especially in morally ambiguous situations. Ultimately, this erosion could potentially undermine managers' own sense of responsibility and clarity about ethical priorities. Managers might be confused about who to trust---for example, their human subordinates, adversarial AI Assistants, or their own judgment, when these perspectives clash. Also, if employees begin to rely on the AI Assistant to voice objections they are afraid to raise issues themselves, they may become deskilled at moral deliberation, but also less confident in being critical on their own, and less trustful in their own perspective.
\end{itemize}

\subsubsection{Employee-Employee Relationship:}
An adversarial AI Assistant could encourage critical evaluation of ideas among peers to prevent groupthink, while also fostering a collaborative team environment. Thus, adversarial AI assistants could improve group performance by enhancing creativity and avoiding groupthink. In fact, it has been shown that groups perform better when someone plays the ``devil’s advocate''\cite{21,22,23,24}. Similarly, an adversarial AI Assistant could push colleagues to question assumptions, identify overlooked risks, and strengthen the quality of proposals without fundamentally eroding mutual respect and teamwork.

\begin{itemize}
    \item\textbf{Example:} In a marketing team, an adversarial AI Assistant could encourage one employee to question a popular but ethically dubious campaign idea (``this strategy could reinforce harmful stereotypes against women, should we rethink it?''). Another example: In engineering, an adversarial AI Assistant could challenge a colleague’s quick fix to a bug by pushing consideration of long-term scalability and user impact.
    \item\textbf{Ethical Issues:} Persistent adversarial feedback could erode trust and goodwill between colleagues, undermining team cohesion, which might escalate small disagreements into larger conflicts, harming morale and collaboration. Also, employees may become hesitant to propose new ideas if they fear constant AI-supported critique from peers. More broadly, if the adversarial AI Assistant become too antagonistic, there is s a risk of eroding the firm's culture that should support effective teamwork, undermining the firm’s stated values and damaging its internal integrity.
\end{itemize}

\subsection{Diverse}

Diverse AI Assistants are designed to embed and present multiple (and potentially conflicting) perspectives on a given issue. This approach aims to help both managers and employees navigate trade-offs between competing values by offering a glimpse into different ethical frameworks and stakeholder interests. Rather than reinforcing a single viewpoint for the firm, a diverse AI Assistant would aim to foster the value of pluralism, encouraging users to appreciate the complexity of ethical decision-making. A diverse AI Assistant would be fine-tuned on datasets where the same issue is addressed from various ethically legitimate perspectives (a product discussed from the standpoint of user benefit, environmental sustainability, investors' interests, and employee well-being) but also from different ethical frameworks (deontological, utilitarian, virtue ethics, etc.). In a way, diverse AI Assistant could operationalize the Rawlsian idea of reflective equilibrium within firms---assessing general principles against cases and revising case judgments in line with general principles so as to converge on a robust set of principles.   

\subsubsection{Manager-Employee Relationship:}

The purpose of a diverse AI Assistant would be to help managers recognize competing interests of different stakeholders (employees, shareholders, regulators, customers) by offering different types of arguments aligned with those stakeholders’ interests. The goal is not to push managers toward or against company goals (as with aligned or adversarial models), but rather to help broaden the ethical landscape of a decision. For employees, a diverse AI Assistant could serve as a counselor in situations where they are caught between duties towards their managers and other professional or ethical obligations.

\begin{itemize}
    \item\textbf{Example:} When deciding how to allocate budget, a diverse AI Assistant might present a manager with multiple viewpoints such as: (1) cost-cutting to maximize shareholder returns; (2) enhancing employee well-being by allocating more funds for training and support; (3) prioritizing customer satisfaction by investing in user research and quality assurance; and (4) environmental concerns. So the manager could have a more exhaustive view/more information regarding the ethical implications of their decision, helping them make a more informed decision. For an employee caught between a manager’s expectations and other duties, an AI Assistant could surface multiple viewpoints: one aligned with managerial expectations, another highlighting collegial duties, and a third reflecting professional or ethical standards. The employee could then decide on their own which one is better.
    \item\textbf{Ethical Issues:} Managers may struggle to choose between equally valid but conflicting options, delaying action. Also, continual exposure to diverging stakeholder concerns could erode a manager’s ability to provide clear guidance and exercise their leadership goals. Managers may also blame the complexity of options for poor outcomes, diluting personal responsibility for final choices. For employees, these AI Assistants may also introduce ambiguity and decision fatigue, especially when the AI Assistant presents multiple versions and viewpoints, without guidance on how to weigh them. This can place undue pressure on employees to navigate ethical tensions without clear institutional backing.
\end{itemize}
  
\subsubsection{Employee-Employee relationship:}

A diverse AI Assistant could foster mutual understanding and dialogue among employees from different backgrounds and departments by exposing them to the ethical and strategic merits of different working styles and priorities. Instead of pushing employees to align on a single corporate value (as in aligned AI Assistants) or to challenge each other’s ideas directly (as in adversarial AI Assistants), diverse AI Assistants showcase multiple perspectives, encouraging creativity and broader value sensitivity within teams. This can help teams navigate complex decisions where no single perspective is obviously correct, and where progress requires integrating multiple views into workable, respectful compromises. 

\begin{itemize}
    \item\textbf{Example:} In an advertisement company, a diverse AI Assistant might present a designer’s aesthetic interests, a project manager’s cost-efficiency focus, and a copy’s stress on creativity and risk-taking side-by-side, offering ethical and strategic reasoning for each. This would allow colleagues to appreciate one another’s interests and weigh trade-offs, instead of competing for dominance of their own perspective.
    \item\textbf{Ethical Issues:} While intended to promote understanding, diverse AI Assistants can backfire if not embedded within a culture that values openness and honesty. Employees might select the AI outputs that confirm their pre-existing biases rather than engaging with the full range of perspectives. Divergent AI-backed perspectives could increase friction rather than mutual understanding if employees use AI outputs to justify opposing positions. Also, employees might be overwhelmed by nuanced trade-offs without clear resolution mechanisms, slowing down collaboration and decision-making. Without preexisting shared decision-making norms or a strong collaborative culture, diverse AI Assistants may slow collaboration and complicate accountability, leaving teams unsure of whose judgment ultimately prevails.
\end{itemize}

\section{Conclusion}
\label{sec:conclusion}

There is no one-size-fits all strategy for aligning AI assistants deployed in business contexts. Rather, each strategy fits a different firm's profile and mission. Supportive AI Assistants might be better suited in well-established companies, which already have an efficient workflow routine and strong ethical governance. Adversarial AI Assistants, on the other hand, might be more fitted for companies that are more interested in growth and change and where thinking outside-the-box is the main driver for these. And diverse AI Assistants are suitable for companies where decisions and their stakes are high but also ambiguous because stakeholder values legitimately diverge, and no single ethical framework yields an answer that holds in all cases. 

What is crucial is that we recognize that AI Assistants have perspectives that will impact significantly the the firms in which they are deployed. Each AI Assistant comes with a perspective which affects how employees within firms think, act, and relate to one another. Accordingly, leaders and decision makers in firms must be intentional in how they select, develop, and deploy these AI Assistants so as to retain control over the cultural and moral fabric of their firms. 

\section*{Ethical Statement}
This research did not involve human subjects and did not require IRB review. Nonetheless, we recognize that our affiliations with commercial and academic institutions involved in deploying or researching LLMs may influence how we frame risks and solutions. We have therefore aimed to critically assess, as opposed to simply promote, the use of alignment strategies within firms. We emphasize the need for caution, transparency, and contextual awareness in their deployment.

\section*{Adverse Impact Statement}
There are risks associated with the use and interpretation of our proposed alignment strategies. If misapplied, these strategies may reinforce dominant norms, encourage over-reliance on AI systems, or marginalize dissenting perspectives within firms. Moreover, firms could adopt alignment strategies selectively to support pre-existing agendas rather than to foster ethical reflection. Our analysis is meant to highlight such risks and encourage intentional, pluralistic approaches to AI deployment that account for intra-firm relational and ethical complexity.

\bibliography{aaai25}

\end{document}